# First Results from the DRIFT-IIa Dark Matter Detector

S. Burgos [a], J. Forbes [a], C. Ghag [b], M. Gold [c], V.A. Kudryavtsev [d], T.B. Lawson[d], D. Loomba [c], P. Majewski [d], D. Muna[d], A.StJ. Murphy [b], G.G. Nicklin [d], S.M. Paling [d], A. Petkov [a], S.J.S. Plank [b], M. Robinson [d], N. Sanghi [c], N.J.T. Smith [e], D.P. Snowden-Ifft [a,*], N.J.C. Spooner [d], T.J. Sumner [f], J. Turk [c], E. Tziaferi [d]

*Corresponding author.

*E-mail address*: ifft@oxy.edu (D.P. Snowden-Ifft)

[a] *Department of Physics, Occidental College, Los Angeles, CA 90041, USA.*

[b] *School of Physics, University of Edinburgh, Edinburgh, EH9 3JZ, UK.*

[c] *Department of Physics and Astronomy, University of New Mexico, NM 87131, USA.*

[d] *Department of Physics and Astronomy, University of Sheffield, Sheffield, S3 7RH, UK.*

[e] *Particle Physics Department, Rutherford Appleton Laboratory, Chilton, UK*

[f] *Blackett Laboratory, Imperial College London, UK*

**Abstract**

Data from the DRIFT-IIa directional dark matter experiment are presented, collected during a near continuous 6 month running period. A detailed calibration analysis comparing data from gamma-ray, x-ray and neutron sources to a GEANT4 Monte Carlo simulations reveals an efficiency for detection of neutron induced recoils of 94±2(stat.)±5(sys.)%. Software-based cuts, designed to remove non-nuclear recoil events, are shown to reject $^{60}$Co gamma-rays with a rejection factor of better than $8\times10^{-6}$ for all energies above threshold. An unexpected event population has been discovered and is shown here to be due to the alpha-decay of $^{222}$Rn daughter nuclei that have attached to the central cathode. A limit on the flux of neutrons in the Boulby Underground Laboratory is derived from analysis of unshielded and shielded data.

## 1. Introduction

The determination of the nature of dark matter remains one of the outstanding goals of contemporary physics. The presently favored model for galaxies, including the Milky Way, posit the existence of a galactic halo composed of Weakly Interacting Massive Particles (WIMPs) [1] which could be the lightest supersymmetric particle (LSP) suggested by extensions of the Standard Model of particle physics [2]. Confirmation of this model requires the direct observation of WIMP-induced nuclear recoils and to this end several experiments have been developed using various technologies. These include the NAIAD and ZEPLIN experiments at the Boulby Underground Laboratory in the UK [3][4], CDMS at the Soudan mine in the US [5], and Edelweiss and XENON in the Gran Sasso in Italy [6,7].

While these experiments are capable of identifying nuclear recoils, only the DRIFT project utilizes a technology that can take advantage of the predicted WIMP directional signature [8]. This arises from the expectation that the galaxy revolves within an effectively stationary WIMP halo such that an Earth-bound detector sees a WIMP 'wind' incident from the direction of the Solar System's motion through the galaxy. Measurement of the resulting non-isotropic recoil distribution, with respect to a sidereal reference frame, produces an asymmetry many times larger than that of annual modulation [8]. Furthermore, there is the potential to perform WIMP astronomy to determine the true kinematics of particles in the Galactic halo [9].

The DRIFT-IIa instrument developed out of earlier work [10, 11, 12, 13] that explored the use of a negative ion time projection chambers (NITPC) filled with carbon-disulfide ($CS_2$) gas. Energy deposition resulting from an interaction of an incident particle with a target gas nucleus produces electron−ion pairs that are then separated by a strong electric field applied between a central cathode and readout planes. $CS_2$ is electronegative, and thus it is negative ions, rather than electrons, that are drifted through the detector. These anions experience only thermal diffusion allowing 3D reconstruction of the recoil track range and direction [8] without the use of a magnetic field. On arrival at one of the multi-wire proportional chambers readouts (MWPCs) the anions are stripped of their electrons; these then avalanche in the normal way to produce signals on the anode and grid wires. The NITPC technology developed for DRIFT can also be used to search for Kaluza-Klein axions through observation of their decay to two gamma-rays [14].

DRIFT-IIa is the first module of a test array of second-generation 1 $m^3$ low-pressure NITPCs and is located in the Boulby Underground Laboratory at a depth of 1070 meters (2805 meter water equivalent) [15]. The present paper focuses on neutron detection efficiency, the gamma-ray discrimination power, a study of the effects of radon and a measurement of the flux of neutrons in the Boulby Underground Laboratory. Work on the directional sensitivity and response to alphas of DRIFT-IIa will be covered in detail in subsequent papers.

**2. The DRIFT-IIa detector**

A full description of DRIFT-IIa is given in [16]; a short summary is provided here for convenience. A $1.5^3$ $m^3$ low background stainless steel vacuum vessel provided containment for 40 Torr of $CS_2$. The $CS_2$ gas, supplied by evaporation from the liquid state, flowed through the vessel steadily by means of a mass flow controller at the input and a dry pump at the output. The flow through the system was equivalent to a 1/e flush time of 3.15 days. Within the vacuum vessel were housed two back-to-back TPCs with a shared, vertical, central cathode constructed of 20 μm stainless steel wires with 2 mm pitch, see Fig. 1. Two field cages, located on either side of the central cathode, defined two drift regions (Z-direction) of 50 cm depth in which recoil events could form. Charge readout of tracks was provided by two MWPCs each comprised of an anode plane of 20 μm stainless steel wires with 2 mm pitch sandwiched between (1 cm gap) two perpendicular grid planes of 100 μm stainless steel wires also with 2 mm pitch. The electrical potential between the grids and the grounded anode planes was -2984 V, which yielded a gas gain of ~1000. The central cathode voltage at -34 kV produced a mean operating drift field of 624±4 V/cm. For each MWPC, 512 grid wires (Y-direction) were grouped down to 32 channels, passed to 32 preamplifiers, the outputs of which were further grouped down to 8 lines (henceforth 'lines' is used to indicate a grouped readout of wires) for shaping, further amplification and digitization. The anode wire signals (X-direction) were more simply grouped to 8 lines which were then pre-amplified, shaped and digitized. Eight adjacent readout lines (either grid or anode) therefore sampled a distance of 16 mm sufficient to contain all candidate WIMP recoil tracks [13]. Voltages on the grid lines were sampled at 2 MHz while the anode lines were sampled at 1 MHz. The 20 wires at the edges of the grid and anode planes were grouped together to provide veto signals for each MWPC. All lines were digitized from -1000 μS to +4000 μS relative to the trigger with 12 bit digitizers. Triggering of the data acquisition system (DAQ) occurred if the sum of the grid lines fell below -100 mV (roughly -200 ADC counts). As discussed in [10], the region bounded by the vetoes and the inner grid planes formed a fiducial volume of 1 $m^3$ equating to 167 g mass of $CS_2$ within which recoil events could occur. Each side of the detector was instrumented with an automated, retractable, ~100 μCi $^{55}$Fe calibration source which allowed periodic (every 6 hours) monitoring of detector gain and functionality.

## 3. Experimental configurations

DRIFT-IIa was placed within the scientific laboratory of the Boulby Mine and data accumulated in both unshielded and shielded configurations. Here we describe these configurations and present estimates of the ambient neutron flux incident upon the detector in both cases. A measurement of the actual ambient neutron flux follows in section 8.

Below the DRIFT-IIa detector, a cavity 10 m long, 3.3 m wide and 45-55 cm deep was constructed providing sufficient space for under-floor passive neutron shielding. Regularly spaced (10 cm wide, 25 cm deep) wooden beams laid over central 1 m wide columns of ~15 cm polypropylene sheet provide structural support to the floor. The remaining gaps are filled with loose polypropylene pellets. Above this floor three adjustable steel legs raised the base of the DRIFT-IIa vessel ~25 cm. The resulting gaps were, again, filled with loose polypropylene pellets. Overall, the structure provided an average of 44 g cm$^{-2}$ of hydrogenous shielding beneath the detector with no significant areas of less than 35 g cm$^{-2}$. In the shielded configuration, additional shielding was constructed such that DRIFT-IIa was completely enclosed within a layer of polypropylene pellets, of typical thickness 44 g cm$^{-2}$.

The expected unshielded rate was calculated as follows. The entire scientific laboratory at the Boulby mine, including all shielding, was simulated using a GEANT4 [17] Monte Carlo. The rock surrounding the cavern was simulated as pure NaCl of 2.2 g cm$^{-3}$ density with U and Th contamination of 60 ppb and 130 ppb, respectively. These values were determined by open Ge detector measurements of gamma-ray background in the laboratory where DRIFT-IIa was located [18]. They are also consistent with a recent measurement of the neutron flux in another area of the Boulby mine [19]. A modified version of the SOURCES code was used to calculate the production rate and spectrum of neutrons entering the laboratory due to the decay of radioactive isotopes in the decay chains of U & Th for input to the GEANT4 simulation [20]. This gave a neutron production rate within the rock of $6.285\times10^{-8}$ neutrons cm$^{-3}$ s$^{-1}$ with a mean energy of 1.73 MeV. Within the simulation these neutrons were produced isotropically from the rock in a volume up to 3 m deep from the cavern-rock face giving a total active rock volume of 7836 m$^3$, a rock surface area of 4085 m$^2$, and a neutron production rate for this volume of 493 neutrons s$^{-1}$. That a depth of 3 m would be sufficient to accurately predict the flux emanating from the walls was determined via an independent simulation which indicated that >99.9% of the neutrons that escape the rock into the cavern originate from within a 3 m depth of rock.

From the simulations it was found that the flux of neutrons at the rock-laboratory boundary was $4.53\times10^{-6}$ cm$^{-2}$ s$^{-1}$, while the flux falling on the outer surfaces of the DRIFT-IIa vessel was found to be $7.19\times10^{-6}$ cm$^{-2}$ s$^{-1}$. This apparent increase in flux was identified as being due to neutrons scattering in the walls of the cavern and undergoing multiple traversals of the laboratory. The simulated neutron recoil rate within the fiducial volume of the DRIFT-IIa detector is shown in Table 1. To reduce the computational burden the density of the CS$_2$ gas was simulated as being 25 times higher than in the actual experiment; other simulations confirmed that this would have no impact on the results of the simulation.

The impact of passive neutron shielding has also been investigated. This work suggests that enclosing the vessel within a layer of ~40 g cm$^{-2}$ of CH$_2$ reduces the rate of nuclear recoils within by more than a factor of 300 which suggests a rate of background neutron events within the shielded configuration of DRIFTIIa of approximately 1 per year. This is in agreement with previous work [20].

The data reported here come from 6 kg days in which the detector was not enclosed within neutron shielding and 10.2 kg days in which the detector was fully enclosed within neutron shielding. Neutron ($^{252}$Cf) and gamma ray ($^{60}$Co) calibration data sets were obtained in both configurations. Remote slow control and data acquisition systems [10] enabled continuous unmanned operation, interrupted only to replenish CS$_2$, retrieve data and perform certain calibration runs. Logged slow control parameters show the detector performance throughout to have remained within expected specifications.

## 4. Data

Some examples of events can be found in Figures 2-4. For each event only a time slice around the event trigger is presented. The left and right plots show data taken from the left and right MWPCs respectively. As discussed bove, the 512 anode and grid wires were grouped down to 8 lines for digitization. The 8 digitized lines from the grid are shown at the top while the 8 digitized lines from the anode are shown to the bottom. The waveform presented below each set of 8 lines is the 'sum line', calculated simply by summing the 8 individual contributions. Two features should be noted. Firstly, in reality the voltages on the wires for the grid and the anode are of opposite polarity but, for ease of viewing, one of them has been inverted; by convention ionization falling on the anode is shown with a negative polarity. Secondly, the gain of the amplifiers for anode signals was lower by a factor of ~2.5 than the gain of the amplifiers for the grid.

Shown between the anode and grid lines are the veto lines. The anode veto line is shown in light blue while the grid veto line is shown in grey. In light of the convention discussed above, an event on these lines caused by, for instance, the electrical ground level changing, would appear to have opposite polarities on the grid and anode veto lines and thus the signal seen on the sum line, shown in black, would be zero. Conversely, ionization falling on only the grid veto or anode veto, or both, would appear on the summed veto line as a negative going pulse.

The left and right vertical grey lines show the region of interest (ROI) for that event while the central grey line shows the trigger time. The horizontal grey lines below the waveforms show the baselines (mostly hidden) and software thresholds for those lines. The vertical black hash marks on individual lines show the start and stop times for lines surpassing the threshold.

The waveforms presented in Figures 2-4 clearly show differing morphologies, each of which may be identified with a particular class of event type. The task of the event analysis was to implement automated software routines to identify with high confidence and high efficiency the waveforms that indicate nuclear recoil events, and for these events to determine the relevant physical parameters, such as ionization, track length and orientation of the recoiling nucleus. The scope of this paper does not include a detailed discussion of this analysis. Much of the analysis procedure has already been discussed in [13]. A full description of analysis of DRIFT-IIa data can be found in found in [21].

## 5. Fe calibrations

Work by this collaboration to be published elsewhere [21] has shown that a parameter $\Sigma$ may be constructed that is proportional to the integral of the voltage and therefore proportional to the charge falling on the anode wires. In order to monitor this proportionality constant, each side of the detector was exposed every 6 hours to a $^{55}$Fe calibration source. The centroid of the $^{55}$Fe photo-absorption peak at 5.9 keV was measured and corresponds to 300±40 number of ion pairs (*NIPs*) at 40 Torr in $CS_2$. Gain variations (as much as 50%) were measured to ~2% accuracy, providing time-dependent calibration constants. Again full description of this analysis data can be found in found in [21].

## 6. Nuclear recoil trigger efficiency

Critical to the success of any dark matter experiment is an understanding of its response to nuclear recoils. To this end, the DRIFT-IIa detector was exposed to a $^{252}$Cf source in a variety of geometries over an extended period of time. Neutrons from the $^{252}$Cf source, interacting with the DRIFT-IIa gas, produced S recoils well matched to those expected from WIMP interactions. C recoil events and Compton scattering events, produced by the gamma-rays emitted from the $^{252}$Cf source, were also, of course, produced. The activity of the source at the time of the exposures was 11600±600 neutrons/sec (manufacturer's information and an independent measurement). This source was contained within a lead canister of wall thickness 1.3 cm and outer dimensions 5 cm diameter by 11 cm length. Experimental runs were performed with DRIFT-IIa exposed to this canister directly. When not used the canister was stored in a tackle box filled with wax of outer dimensions 22cm×25cm×35cm when closed. Further exposures were conducted with this tackle box opened, in which case the upper 13 cm of the box was moved sideways, completely exposing the top region of the lead canister. Table 2 summarizes the 5 experimental configurations in which $^{252}$Cf source data were collected.

Fig. 5. shows the trigger rate as a function of time for a continuous run in which the neutron source was present at the beginning of the run and then removed. As can be seen, there is a clear transition in the raw data acquisition event rate corresponding to the removal of the source. In order to compare these data to theory a simple set of cuts were required so that *NIPs* could be accurately calculated for the triggered events. These cuts were developed to allow DRIFT-IIa to be used as an accurate proportional counter. They also serve as the basis for additional cuts to remove background events, discussed later. A full description of the analysis of the DRIFT-IIa data can be found in found in [21]. A short description of these cuts is provided here.

First, all waveforms were required to fall within the limits of the digitizer to ensure that the Σ parameter be accurately measured. Second, no ionization was permitted on the veto lines ensuring that the events were fully contained. Finally, for each triggered event, 1000 μS of pre-trigger information and 4000 μS of post-trigger information were saved to disk. As nuclear recoils generated by WIMP are expected to last no more than a few hundred μS, it was found advantageous to focus the analysis on a ROI within the data record. For the nuclear recoil analysis the ROI extended from -200 μS to +500 μS relative to the trigger time. For most of the statistics generated only data within the ROI were analyzed. Waveforms were required to be below threshold at the edges of the ROI to ensure that the ionization could be properly measured.

For each location and configuration of the $^{252}$Cf source, the difference between average rates (with these simple cuts) with the neutron source present and average rates with the neutron source absent were calculated. The results are shown in column 2 of that DRIFT-IIa can be operated in a mode in which, between the experimental threshold and the maximum reasonable energy for WIMP-induced recoils, ~100% efficiency for recording nuclear recoils can be achieved. Moreover the good agreement between theory (which did not include Compton events in the fiducial volume) and experiment is evidence that the detector is insensitive to Compton events in the fiducial volume. The software cuts employed in this mode are, however, unable to reject all non-nuclear recoil events (Compton events in the MWPC for instance), and thus additional event rejection criteria are required. As will be seen below, these result in excellent background rejection, but also impact the WIMP-recoil identification efficiency.

Table 2. Complete simulations using GEANT4 [17] were performed, modeling each of the experiments described; column 3 of Table 2 presents the predicted rates. As will be demonstrated below, hardware thresholds and software cuts resulted in the fiducial volume of DRIFT-IIa being entirely insensitive to Compton recoils induced by gamma-rays from the $^{252}$Cf source. However, *for this analysis*, the higher electric field strength present in the MWPCs and the simple nature of the cuts meant that DRIFT-IIa was sensitive to photon induced Compton events in the MWPCs. Hence, the simulation also modeled the gamma-ray emission from the $^{252}$Cf source and recorded any event that deposited energy within the MWPC. This rate amounted to ~15% of the nuclear recoil rate. The combination of the simulated fiducial volume nuclear recoil rate and MWPC gamma-ray rate produced a simulated rate accurately modeling the experimentally detected rate. Column 4 of Table 2 shows the ratio of the experimental and simulated rates with all statistical errors included. It is worth emphasizing that the good agreement between calculated and measured rates that is demonstrated in Table 2 is achieved for a range of exposure geometries and intensities, separated by several months. This suggests that the response of DRIFT-IIa to neutrons is well understood and stable, that the activity of the calibration neutron source is correctly known, that the normalization of Σ to *NIPs* and energy to *NIPs*, in the Monte Carlo, is being performed consistently, and that the live-time and fiducial volume are well known. While Table 2 shows the overall efficiency within a window of 1000 – 6000 *NIPs*, Fig. 6 shows the efficiency (experimental rate divided by simulated rate) as a function of *NIPs* for a typical case. The fall off in efficiency above 6000 Nips is due to C recoils which are cut in the data due to their long ranges which meant they were not contained in the ROI. This shows

**7. Nuclear recoil analysis**

The overall trigger rate for the detector was ~1 Hz. Most events were clearly not nuclear recoils and could easily be discarded by suitable cuts. The application of the simple cuts discussed above, for instance, reduced the rate to ~0.25 Hz. The alpha-particle event shown in Fig. 3 was cut by these simple cuts because the ionization extended beyond the ROI. Additional cuts were then chosen that optimized nuclear recoil acceptance efficiency, as measured by exposure to $^{252}$Cf neutrons, while simultaneously reduced background events.

As discussed in [13] a class of events called "ringers" formed a large fraction of all events recorded to disk. Several features of these events, discussed in detail in [21], show that they are not associated with ionization of recoiling electrons or nuclei, most significantly, that their rate did not increase in the presence of gamma-ray or neutron producing sources. Ringer events were therefore removed from the dataset.

Ionization created inside the MWPC, whether caused by sparks or ionizing particles,

was rapidly collected on the anodes due to the large electric field in this region. Cutting such events with a width cut effectively provided a veto for the two remaining sides making up the cube of the DRIFT-IIa fiducial volume.

Nuclear recoil events in the energy range of interest for WIMP searches (~1 keV/amu [22]) are expected to be short, see [13]. Therefore, any event which caused all 8 anode lines to fall below threshold was cut. A C or S recoil that registers on all 8 anode wires is at least 12 mm long as the anode wires are separated by 2 mm. A S recoil with an average range of 12 mm has an energy of 660 keV (~20 keV/amu) and produces 19,000 *NIPs* [13]. A C recoil with an average range of 12 mm has energy of 120 keV (~10 keV/amu) and produces 4,200 *NIPs* [13]. If two or more anode wires were "hit" (voltages go below threshold) they were expected to be adjacent to each other. Finally for some events it was observed that a significant fraction of the ionization appeared on lines which did not trigger the analysis. This is in contrast to the expectation that nuclear recoils should have a high density of ionization and therefore these events were removed from further analysis.

The likelihood that the short nuclear recoil events DRIFT was designed to detect would appear on both sides of the detector (crossing the central cathode) is extremely small. Nevertheless, in the background data (unshielded and shielded) such events were found with an unexpectedly high frequency, see Fig. 7. The origin of these events is identified and discussed later, but for the analysis presented here it was appropriate to implement a cut that removed these events from the data.

Finally, some events were observed in which ionization appeared, unexpectedly, on the anode prior to the start of the ROI. This type of ionization was called pre-ionization.

Events of the same character were observed in the DRIFT-I data. Fig. 8 illustrates an example of such an event from DRIFT-IIa. The likely origin of these events is discussed later, but for the analysis presented here it was appropriate to remove them.

These cuts reduce the accepted rate of events by a factor of ~350 relative to the trigger rate. They also reduce the neutron detection efficiency. This is shown in Fig. 9 which plots the efficiency as a function of *NIPs* for the "Tackle Box" $^{252}$Cf run with the background cuts applied to the data. In the context of Fig. 6 it is clear that neutron events recorded to disk are being removed by these cuts. We believe, however, this set of cuts to be reasonably well optimized. The efficiency curve in Fig. 9 falls faster than that in Fig. 6 at higher *NIPs* values. This, again, is due to long-range C recoils.

**8. Gamma-ray rejection**

A significant source of background in dark matter searches are electrons that result from gamma-ray interactions or, to a lesser extent, radioactive decay. In DRIFT, rejection of electrons relies on their long range and lower ionization density relative to that of a nuclear recoil event [13]. The very low drift speed for ionization in DRIFT-IIa coupled with a ~4 μs shaping time and 1 or 2 MHz readout meant that this low density translated into low amplitude voltage signals on the readout lines. Thus, as discussed in [13] an appropriately set threshold prevents these events from triggering the detector or analysis while still preserving a high trigger efficiency for nuclear recoil events.

In order to measure DRIFT-IIa's gamma-ray rejection capabilities five 0.52 μCi $^{60}$Co sources were placed on top of the unshielded DRIFT-IIa detector, three in the center and two on either side of the central cathode. 0.571 days of live-time data were collected.

A Monte Carlo simulation using GEANT4 [17] was employed that included the running conditions and geometry of the $^{60}$Co exposure. Rates of interaction, energy deposition for each electron recoil, directional information and timing of the events were all simulated. Any event that deposited energy within the fiducial volume was recorded as an interaction and *NIPs* value for that event calculated. The background cuts, including cuts to remove events occurring in the MWPC, discussed above were applied to the data for this analysis and therefore events occurring within the MWPC were not included in the simulation results.

The data analysis technique was similar to that for the calculation of neutron efficiencies, including background cuts as already discussed. A comparison was made to data taken without the sources present, with durations 1.42 and 1.68 live days, immediately before and after the $^{60}$Co data was collected to the $^{60}$Co data. This comparison was complicated by the high rate of Compton interactions during the $^{60}$Co exposure. According to the Monte Carlo, the rate of interactions was 197±1 Hz for a 500 *NIPs* threshold. On average, therefore, there was approximately one 500 *NIPs* or greater Compton event in a 5 ms data record. This extra ionization frequently triggered the cuts associated with ionization falling outside of the ROI causing the observed rate with the $^{60}$Co source nearby, with all background cuts, to be slightly *below* the background rate. These cuts therefore had to be turned off for this analysis. The limits derived below are, therefore, somewhat conservative.

The rate of events surviving analysis cuts without the $^{60}$Co sources present was subtracted from the rate with the sources present. The results of this calculation, as a function of *NIPs*, are shown in second column of Table 3. All of the rate difference data are consistent with zero. The Monte Carlo simulated data are shown in the third column.

The 90% C.L. rejection factors are shown in the 4th column.

## 9. Results and discussion

Fig. 10 shows the distribution of *NIPs* for all retained events accrued in 16.8 days live time of unshielded running. The most striking feature of this histogram is the large peak between ~1000 and ~2000 *NIPs*. The lower limit is a consequence of the hardware trigger threshold and is discussed in detail in [21].

A GEANT4 [17] simulation of the gamma-ray background in DRIFT-IIa was performed to simulate the gamma-ray activity inside an unshielded DRIFT-IIa detector. This calculation revealed that there were $7 \times 10^5$ interactions per day occurring in DRIFT-IIa in an energy range of 1000 – 10000 NIPs. With the appropriate rejection factor above, this implies a background rate of < 2.1 events per day (90% C.L.). The events seen with the unshielded DRIFT-IIa detector were not due to gamma-ray interactions.

The vast majority of the events in Fig. 10 have now been identified as being due to heavy ions recoiling into the active volume of the vessel following alpha-particle decay that has occurred on the wires of the central cathode. Nuclei unstable to alpha-particle decay are located on these wires as a consequence of the presence of Rn, which it is believed has emanated from components within the vessel. $^{222}$Rn is unstable to alpha-particle decay ($t_{1/2}$=3.8 d): when this decays the daughter, $^{218}$Po, will appear as a charged ion and will be attracted to the large negative potential of the cathode [23] where it plates out. $^{218}$Po has a half-life against alpha-particle decay of 3.05 m. Given the random orientation of decay, and that it was occurring on a surface, one expects that some fraction of the time the alpha-particle would have been emitted into the 20 μm diameter central cathode wires. The range of these alpha-particles, in the stainless steel cathode wire, is typically 12-18 μm [24], significantly less than the width of the wires. A simple Monte Carlo of this process suggests that this would have occurred 37% of the time. In these cases the recoiling heavy ion is oriented into the active volume of the detector, leading to a nuclear recoil event being observed. The energy of these events is determined by the fraction of kinetic energy carried off by the heavy ion for the given Q-value of the alpha-decay. In the case of $^{218}$Po, the Q-value for alpha-decay is 6.115 MeV, which results in a recoiling $^{214}$Pb with a kinetic energy of 112.2 keV. The resulting ionization generated is calculated as follows: In [25] it was found that the W factor for $^{214}$Po was 4.5 times higher than the W factor for alpha-particles in Ar. Taking 26.3 eV as the W factor for alpha-particles in Ar [26], multiplying by 4.5 to account for the difference in W factor, and multiplying this by 19.7 eV/26.3 eV to account for the lower W factor in $CS_2$ (300±40 *NIP*s at 5.9 keV as discussed above), and assuming that the W factor for $^{214}$Po is the same as for $^{214}$Pb, gives W = 88.6 eV. This implies a typical ionization level of 1300 *NIPs*. In comparison, the peak of the energy distribution seen in Fig. 10 occurs at ~1500 *NIPs,* very close to this prediction. Events of this type are henceforth classed as radon progeny recoils ('RPR's). Other progeny from this decay chain may also produce RPR events with similar energies.

This hypothesis explains the unexpectedly high number of events with simultaneous ionization on both sides of the detector. On the one side is the RPR while on the other is the accompanying alpha particle which did *not* stop in the wire. It also might explain the pre-ionization events as the pre-ionization may be due to x-rays produced as the alpha particle slowed in the wire, which then interact in the gas ahead of the ionization from RPR.

The presence of radon inside the DRIFT-IIa vacuum vessel has been confirmed in several ways. Firstly, events consistent with $^{222}$Rn decays were observed in the DRIFT-IIa data set. An example is shown in Fig. 11. This event traversed the detector's central cathode, and having no detectable veto signal its origin must have been within the gas. The change in the height of the peaks in this event may be interpreted as a measure of the changing rate of energy loss, dE/dx. The wires are evenly spaced and alpha-particle tracks in gas are approximately straight at these energies [24]. With this in mind the Bragg Peak is visible for this track on the right detector at t ~ 300 μs consistent with the ionizing behavior of alpha-particles. Finally 5.486 MeV alpha-particles from the decay of $^{222}$Rn have a range of 339±13 mm in 40 Torr $CS_2$ [24]. The track range in Δx and Δy were measured by counting peaks and multiplying by the 2 mm spacing of the wires; Δz was measured by calculating the beginning and end of the event, on each side, and multiplying by the measured drift speed. This event was then measured to have a total track length of 351±5 mm in good agreement with the prediction.

Finally, the unshielded data set was re-analyzed with software routines that searched for fully contained alpha-particle tracks, that is long tracks, that both crossed the cathode and did not register on the veto lines. Here we term such events as gold plated cathode crossing (GPCC) events. Fig. 12 shows the rate of GPCC events as a function of time for the first ~two weeks of unshielded data. The red line shows the time when evacuation of the vessel stopped and filling with $CS_2$ started. One would expect the rate of GPCC events to increase and then saturate on a timescale given by the half-life of parent nucleus and the flow rate of gas through the detector according to the following formula,

$$R = \frac{\kappa}{\left(1 + \frac{\tau}{\rho}\right)}\left(1 - e^{-t/\tau}e^{-t/\rho}\right) \tag{1}$$

where $R$ is the rate at which cathode crossing decays occur, $\tau$ is the $^{222}$Rn lifetime (5.52 days), $\rho$ is the *1/e* flush time for the system (3.15 days), and $\kappa$ is the constant rate at which these events are supplied. $\kappa$ is the only free parameter in this formula. As can be seen on Fig. 12 the data are well fit with this function. A similar analysis was performed looking at the time dependence of the RPR events themselves; this is shown in Fig. 13 and is again consistent with this function. That the time evolution of the rate these two disparate types, nuclear recoil and alpha, of events are well fit by the same function is, perhaps, the strongest evidence for the RPR hypothesis.

RPR events were present, as expected, in the shielded data. However, these events form a peak (see Fig. 10) whereas Table 1 indicates that the recoils from neutrons will have a rather hard spectrum. Consequently, a window in *NIPs* from 3000-6000 *NIPs* was chosen for this neutron flux measurement in an attempt to maximize signal to noise. The limits of this window were chosen (i) to avoid, as much as possible, the tail of the RPR peak shown in Fig. 10 and (ii) to avoid the loss of efficiency shown in Fig. 9. Within this window the expected rate of recoils was 0.3±0.1 events per day with a 100% efficient DRIFT-IIa detector.

Interestingly, the events in this window, designed to avoid RPR events, showed a similar increase in rate after pump out as the GPCC and RPR events did in Fig. 12 and Fig. 13, though with much lower statistics. Fig. 14 shows the rate as a function of time for events in this 3000-6000 *NIPs* window in the shielded data. As can be seen, there is a significant increase in rate during the first few days after the detector was pumped out well fitted with the same function used to fit the other rate increases. One possibility is that these are RPR events in which the alpha particle, initially directed into the wire, scatters off of the stainless steel and emerges with greatly reduced energy on the same side of the TPC as the RPR event. While simulations of angular straggling suggest this to be feasible, quantitative estimates of this concept are ongoing.

For simplicity, events for the unshielded/shielded measurement were only considered if they occurred at least 1 week after a pump out ensuring that the rates were at their steady state values, see for instance Fig. 13. For the unshielded data, with data analysis that included cuts as discussed above for detection of nuclear recoils, 61 events were found in 12.03 days of live time giving a rate of 5.1±0.6 events per day. For the shielded data, using identical data analysis and cuts, 67 events were found in 11.1 days of live time, giving a rate of 6.0±0.7 events per day. Subtracting the unshielded from the shielded rate leaves -1±1 events per day, consistent with zero. With all of the cuts in place the average efficiency over all of the neutron exposures performed during the shielded and unshielded runs was 32±1% for the 3000-6000 NIP recoil range. Thus the observed difference in rate implies that the actual difference is -3±3 events per day. An upper limit on the rate is <4 events per day (90% C.L.). If not for the unexpected appearance of the RPR events and the mysterious 3000-6000 *NIP* events sensitivity to the predicted rate could have been achieved.

## 10. Conclusion

DRIFT-IIa is a 1 m$^3$ gaseous TPC detector. First results are reported here for its operation in the Boulby laboratory in which 6 kg days of data were collected with the detector open to the laboratory environment, and 10.2 kg days of data were collected with the device completely enclosed within hydrogenous shielding. A gamma-ray rejection capability is measured to a level of better than $8\times10^{-6}$ for all energies above a threshold of 1000 *Nips*, while an ability to record 94±2(statistical)±5(systematic)% of the nuclear recoils occurring inside the detector within a 1000 < *NIPs* < 6000 ionization range is demonstrated. Application of additional cuts that reduce the rate of all non-nuclear events by a factor of ~350 are shown to reduce the nuclear-recoil efficiency to ~60% near threshold. An unexpected population of nuclear recoil events has been observed, and are shown to be due to the decay of the daughter nuclei of $^{222}$Rn which have attached to the central cathode. An upper limit on the ambient neutron flux within the Boulby mine has been measured as <4 neutrons day$^{-1}$, while a rate of 0.3 neutron recoils day$^{-1}$ is predicted. Components contributing $^{222}$Rn to the active volume have been identified; removal of these components is underway as well as additional measures to remove or veto these events.

**Acknowledgments**

We acknowledge the support of the US National Science Foundation (NSF). This material is based upon work supported by the National Science Foundation under Grant Nos. 0300973 and 0600789. Any opinions, findings, and conclusions or recommendations expressed in this material are those of the author(s) and do not necessarily reflect the views of the National Science Foundation. We also acknowledge support from the following agencies: the UK's Engineering and Physical Sciences Research Council and Particle Physics and Astronomy Research Council, Cleveland potash Ltd., and ILIAS contract number RII3-CT-2004-506222.

**Tables**

Table 1– The table below shows the predicted rate of recoils from neutrons emanating from the walls of the Boulby Underground Laboratory above a *NIPs* threshold as shown.

| *NIPs* threshold | Rate (per day) |
|---|---|
| 0 | 25.5±0.6 |
| 20 | 8.4±0.3 |
| 100 | 6.2±0.3 |
| 500 | 2.4±0.2 |
| 1000 | 1.4±0.1 |
| 1500 | 1.1±0.1 |
| 2000 | 0.9±0.1 |
| 2500 | 0.7±0.1 |
| 3000 | 0.5±0.1 |
| 3500 | 0.50±0.08 |
| 4000 | 0.35±0.07 |
| 4500 | 0.33±0.07 |
| 5000 | 0.32±0.07 |
| 5500 | 0.27±0.06 |
| 6000 | 0.19±0.05 |
| 6500 | 0.12±0.04 |
| 7000 | 0.12±0.04 |
| 7500 | 0.07±0.03 |
| 8000 | 0.07±0.03 |
| 8500 | 0.07±0.03 |
| 9000 | 0.07±0.03 |
| 9500 | 0.07±0.03 |
| 10000 | 0.07±0.03 |

Table 2

Results for rates, experimental and simulated, in a range 1000-6000 NIPs for various neutron runs. A brief description of the runs follows.

x-neutrons - $^{252}$Cf source, in lead canister, located 1.44 m away from vessel door in X; central in Y & Z. Vessel un-shielded.

y-neutrons - $^{252}$Cf neutron source, in lead canister, located 1.44m away from vessel roof in Y; central in X and Z. Vessel un-shielded.

Tackle Box - $^{252}$Cf neutron source, in lead canister, placed in open wax-filled tackle box. Source 12 cm below vessel. Vessel un-shielded.

Pre-Shielding neutrons - $^{252}$Cf neutron source in lead canister 20cm from DIIa vessel door in X, central in YZ. Vessel 5/6 shielded (face and 30cm section front roof exposed).

| Run type | Experimental Rate (present - absent) (Hz) | Simulated Rate (Hz) | Experiment / Simulation |
|---|---|---|---|
| X-neutrons June 20 2005 | 0.097±0.004 | 0.109±0.005 | 89±5% |
| Y-neutrons June 20, 2005 | 0.110±0.004 | 0.118±0.004 | 93±6% |
| Tackle Box July 12, 2005 | 0.352±0.005 | 0.365±0.009 | 96±3% |
| Pre-shielding neutrons August 10, 2005 | 0.253±0.010 | 0.262±0.009 | 97±5% |
| Weighted Average | | | 94±2% |

Table 3 – Shows the rejection factor for various *NIPs* "windows". The second column shows the measured, source present minus source absent, rate for a particular range of *NIPs*. All results are consistent with 0 Hz. The third column shows the Monte Carlo rate expected for that *NIPs* window. The fourth column shows the calculated rejection 90% C.L. rejection factor. Roughly speaking the first window 1000-2000 *NIPs* includes the bulk of the RPR peak and therefore generates the worst rejection factor. The third window 3000-10000 *NIPs* includes the large events discussed in the text.

| *NIPs* Window | $^{60}$Co rate minus Background rate (Hz) | Monte Carlo rate (Hz) | Rejection Factor Limits (90% C.L.) |
|---|---|---|---|
| 1000-2000 | $(-2\pm5)\times10^{-4}$ | 77±1 | $< 8\times10^{-6}$ |
| 2000-3000 | $(-1\pm1)\times10^{-4}$ | 24.8±0.5 | $< 5\times10^{-6}$ |
| 3000-10000 | $(2\pm6)\times10^{-5}$ | $27.2\pm0.3$ | $< 3\times10^{-6}$ |
| 1000-10000 | $(1\pm3)\times10^{-4}$ | $125.2\pm0.7$ | $< 3\times10^{-6}$ |
| 1000-6000 | $(1\pm3)\times10^{-4}$ | $128.9\pm0.7$ | $< 3\times10^{-6}$ |

**Figures**

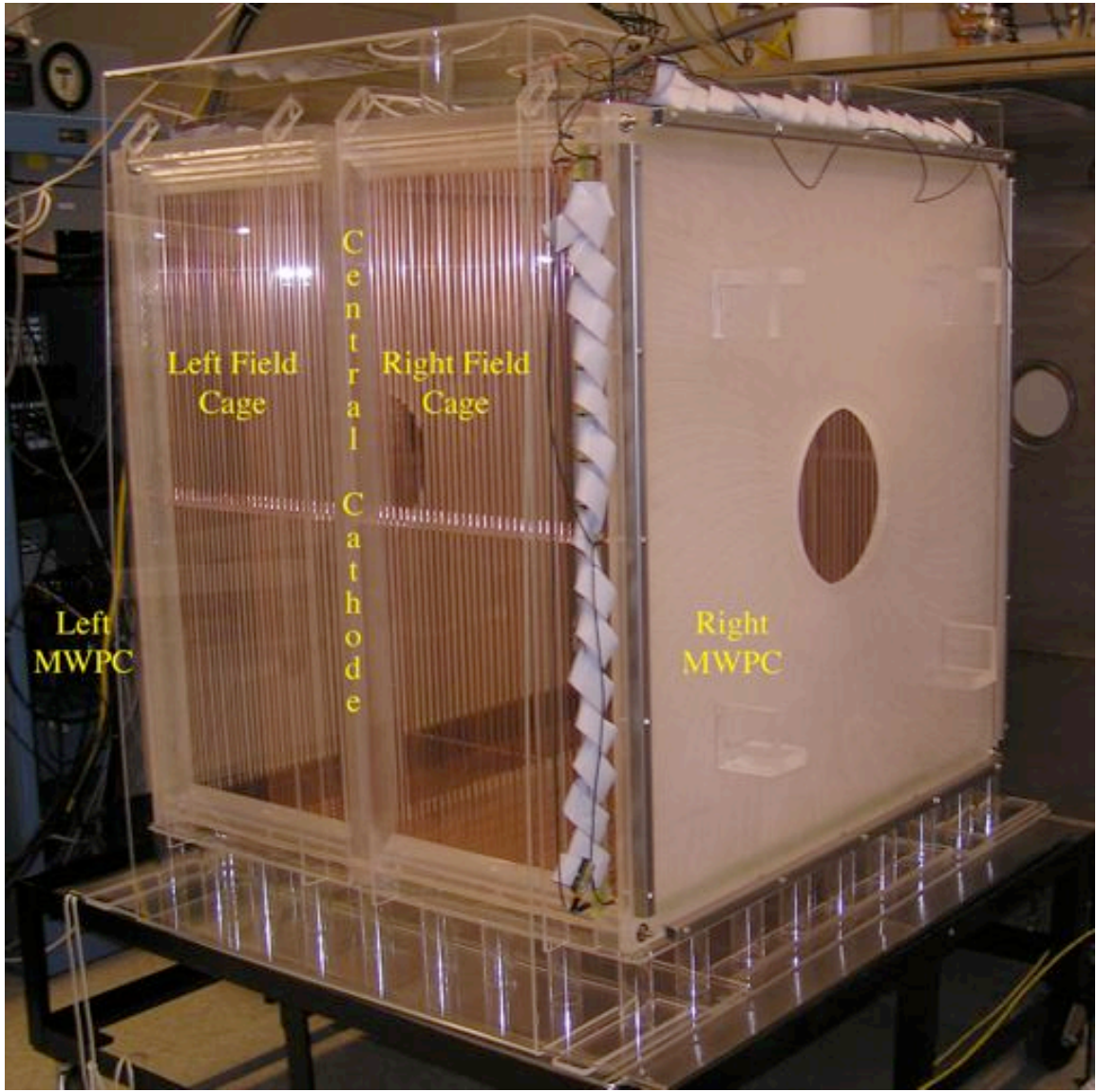

Fig. 1 - A photograph of the DRIFT-IIa detector during commissioning.

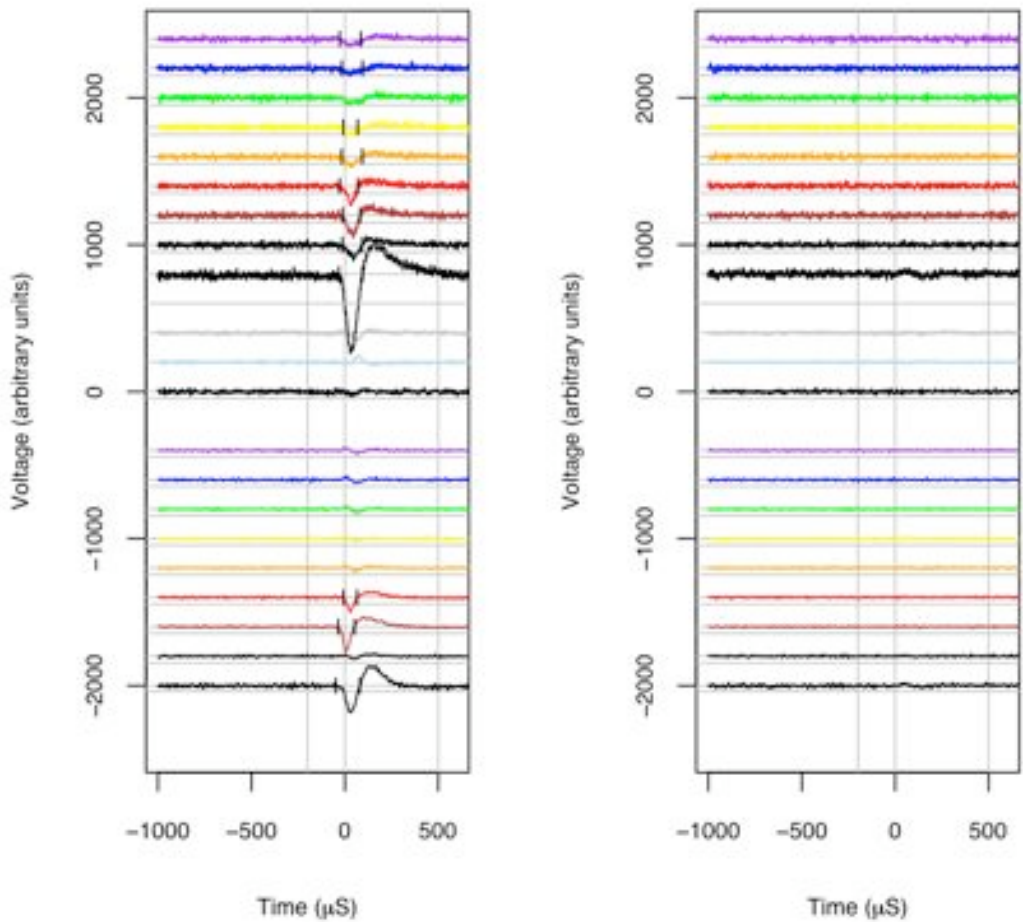

Fig. 2 – Recorded waveforms for an event in which a $^{252}$Cf neutron source was placed near the detector. Background cuts, discussed in the text, reduced the background rate to far less than the observed rate for this data set. Therefore, this was, very likely, a neutron recoil event. The event is clearly visible on lines 2 and 3 (brown and red) on the anode (lower pane) on the left side of the detector. The signals on the remaining anode lines are due to induced electrical pickup and thus appear inverted. The signal on the grid is induced and since the anode-grid separation is quite large (1 cm) the signal on the grid

appears on many wires.

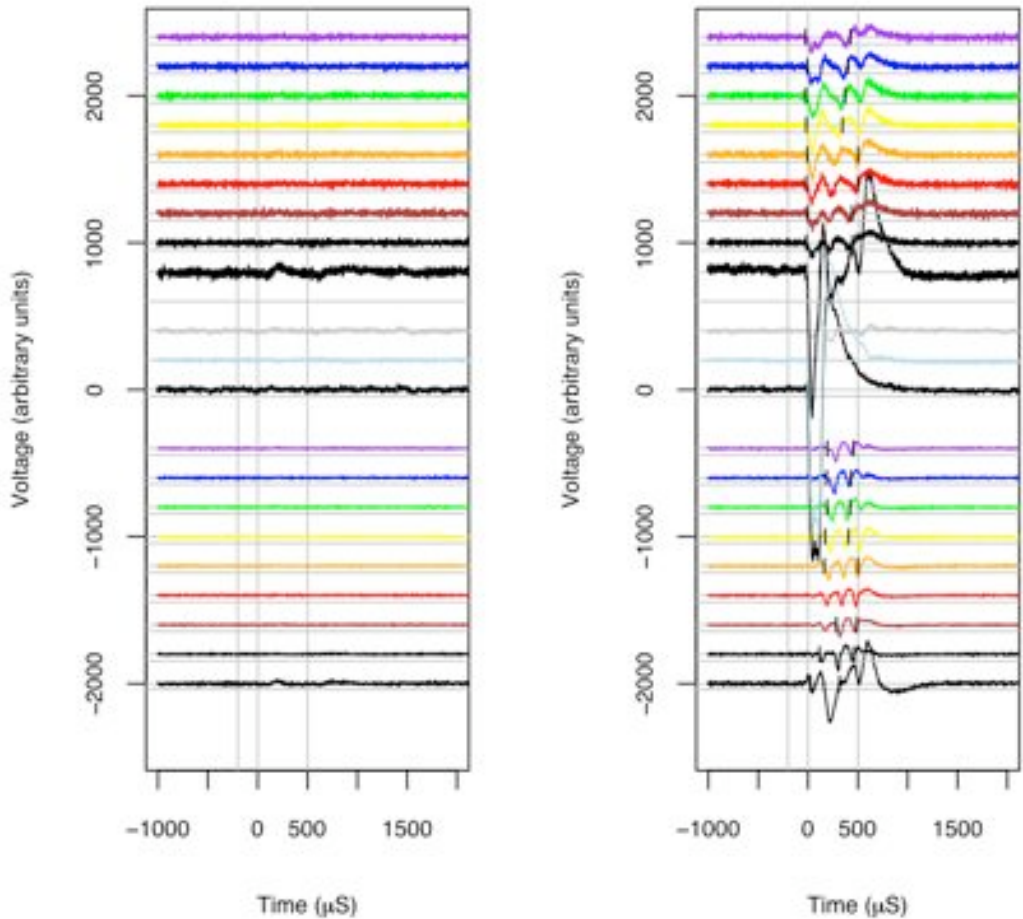

Fig. 3 – A several MeV alpha-particle has a range such that it can travel over many anode and grid wires. If it traverses more than 8 wires it will, because of the wire grouping, reappear later on the same readout line, possibly multiple times. Given the length of this event, the quantity of ionization and the fact that the anode veto line (light blue) received ionization, this is very likely an alpha-particle event.

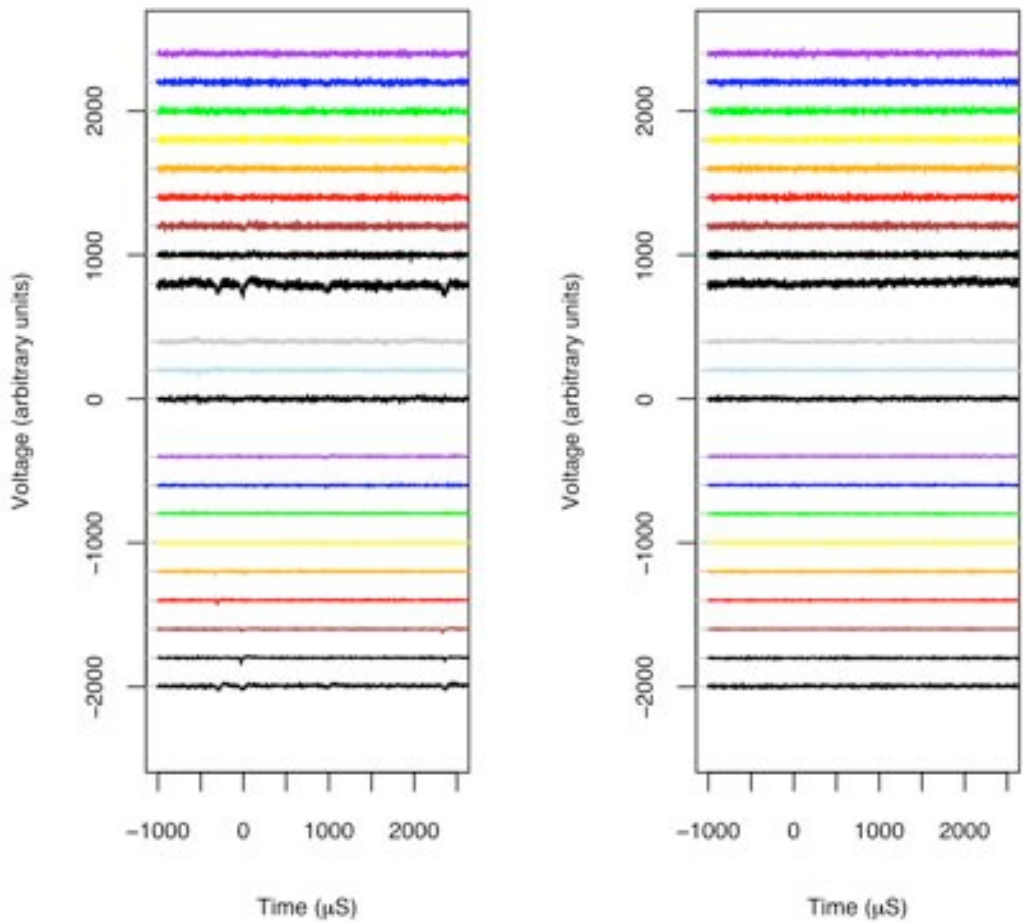

Fig. 4 – As discussed in [10] every 6 hours the left and then the right detector were exposed to an $^{55}$Fe calibration source. In the data shown above the left side was being exposed. 4 events, though small, are visible on the anode and grid. None are visible on the right consistent with the expected attenuation of 5.9 keV x-rays. Because of the way in which these events were analyzed no ROI is shown.

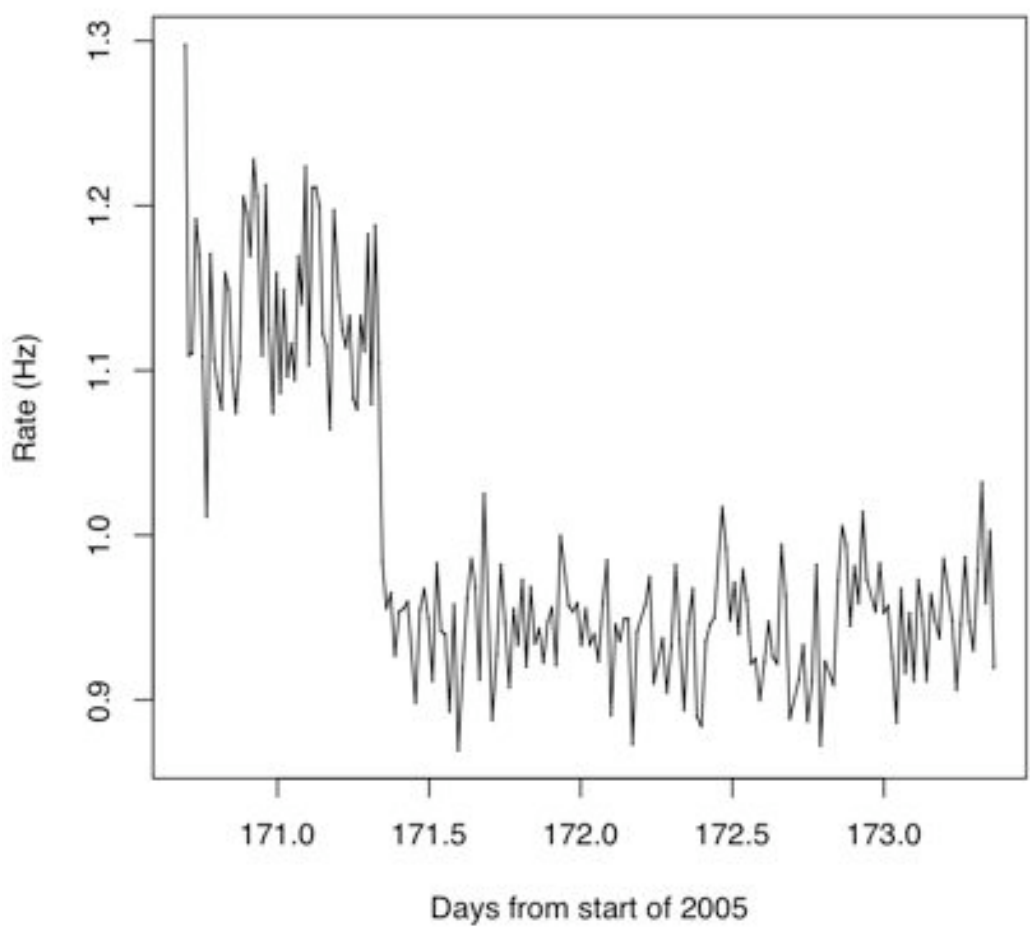

Fig. 5 – This plot shows the trigger rate of the detector for a continuous run. At the beginning of this run the $^{252}$Cf source was 1.44 m away from one face of the vacuum vessel. At a time of ~171.3 days from the start of 2005 the source was removed. There is a clear change in the rate at this time.

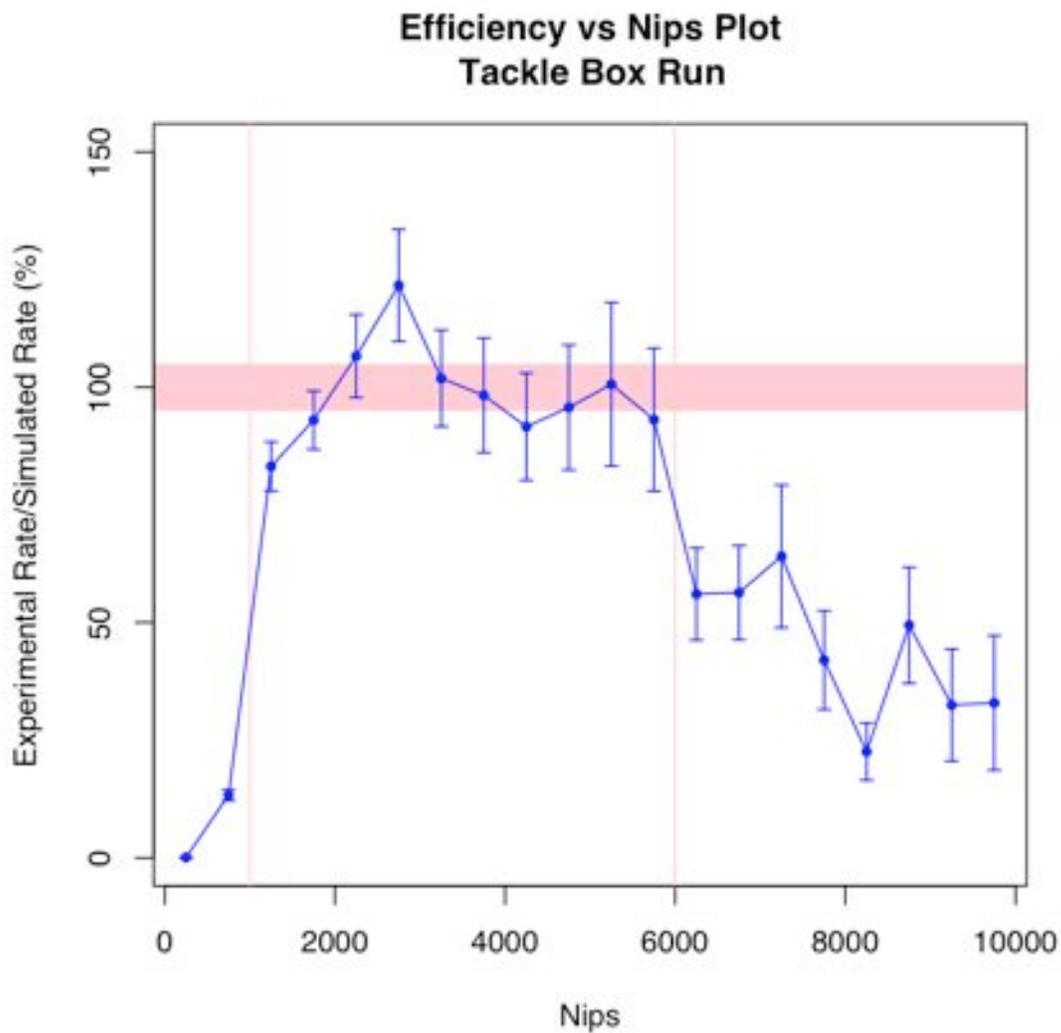

Fig. 6 – The experimental rate divided by the simulated rate as a function of *NIPs* for the "Tackle Box" neutron run. The vertical pink lines indicate the region of interest for the analysis shown in Table 1. The horizontal pink band shows the systematic uncertainty in the source strength, 5%.

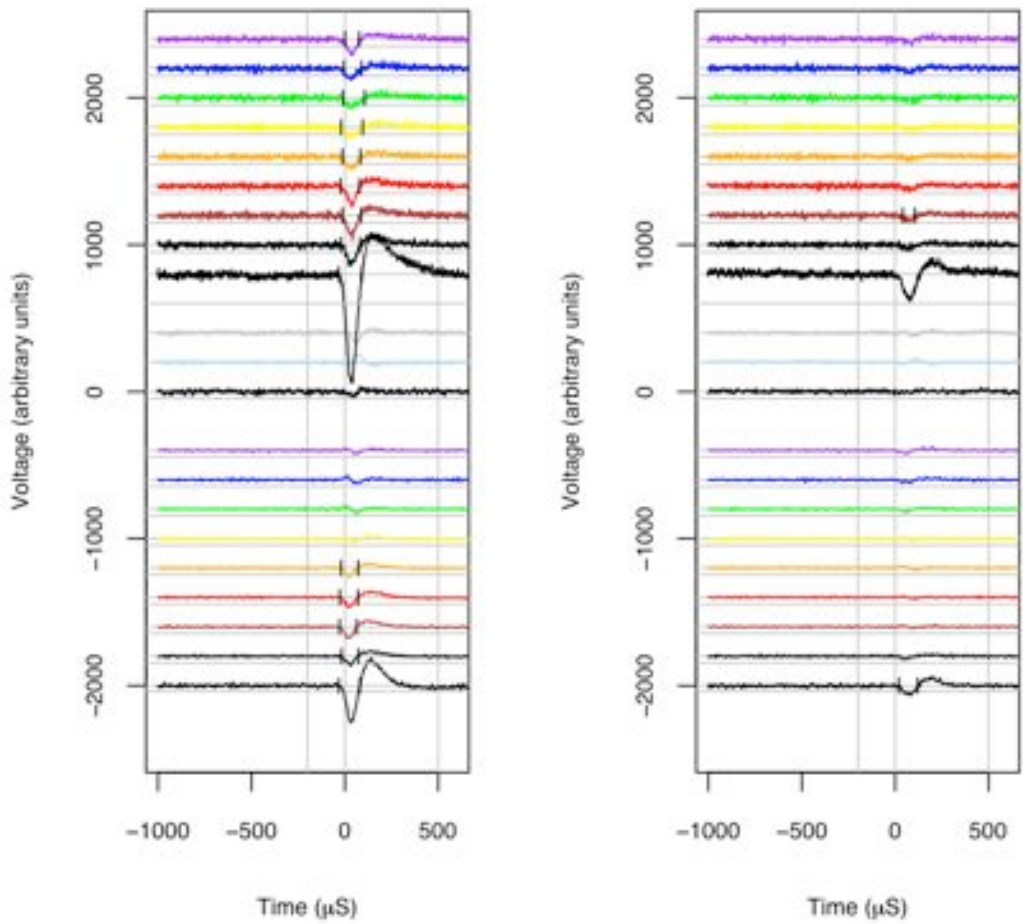

Fig. 7 - A nuclear recoil-like event having simultaneous ionization on both sides of the detector.

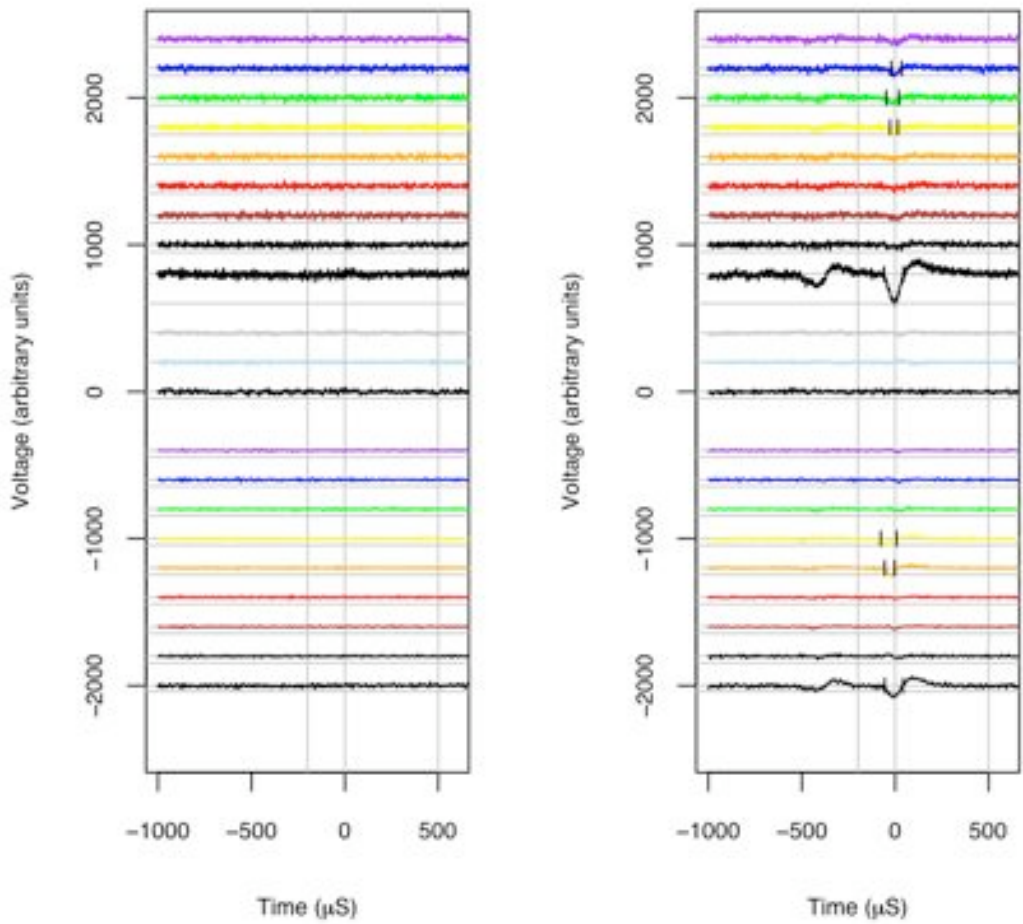

Fig. 8 – An event showing pre-ionization associated with the main event which triggered the system.

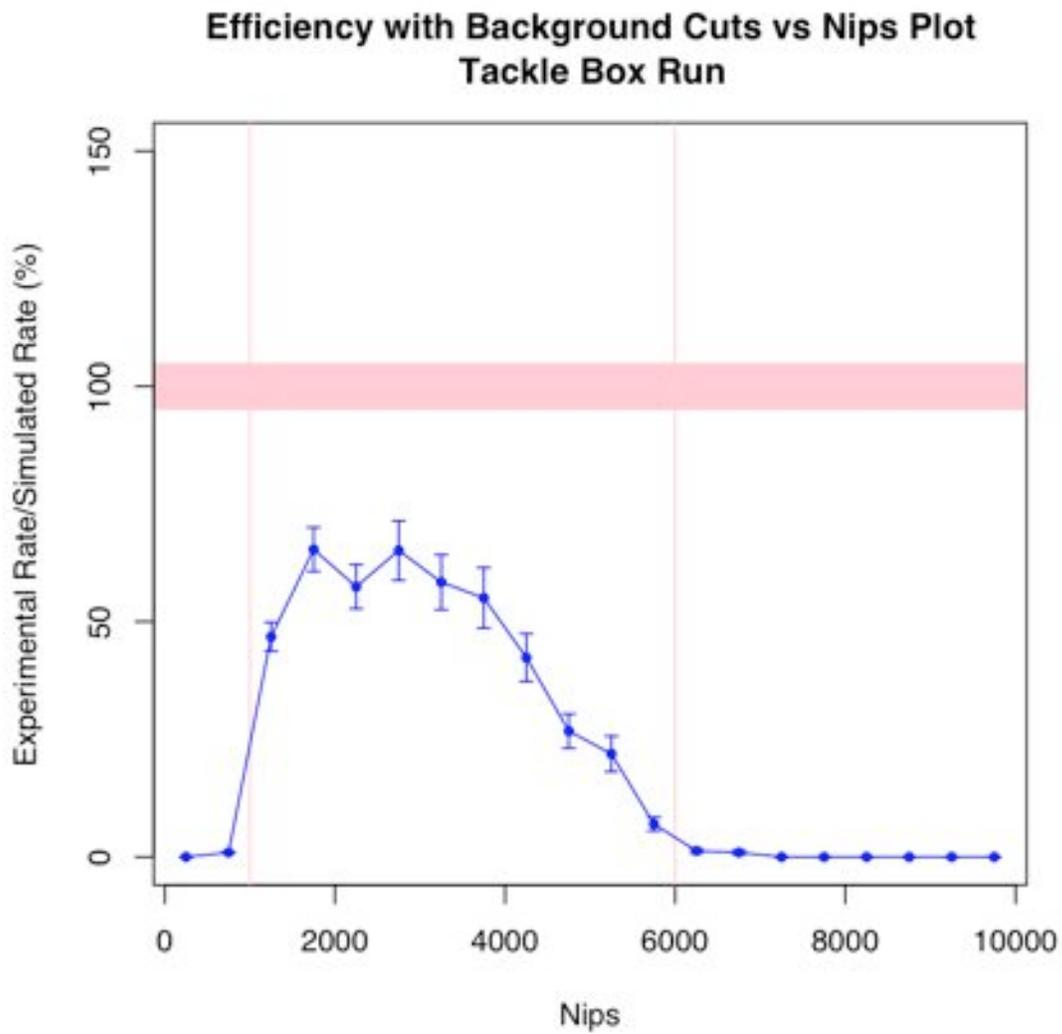

Fig. 9 - The experimental rate divided by the simulated rate as a function of *NIPs* for the "Tackle Box" neutron run with all background cuts in place.

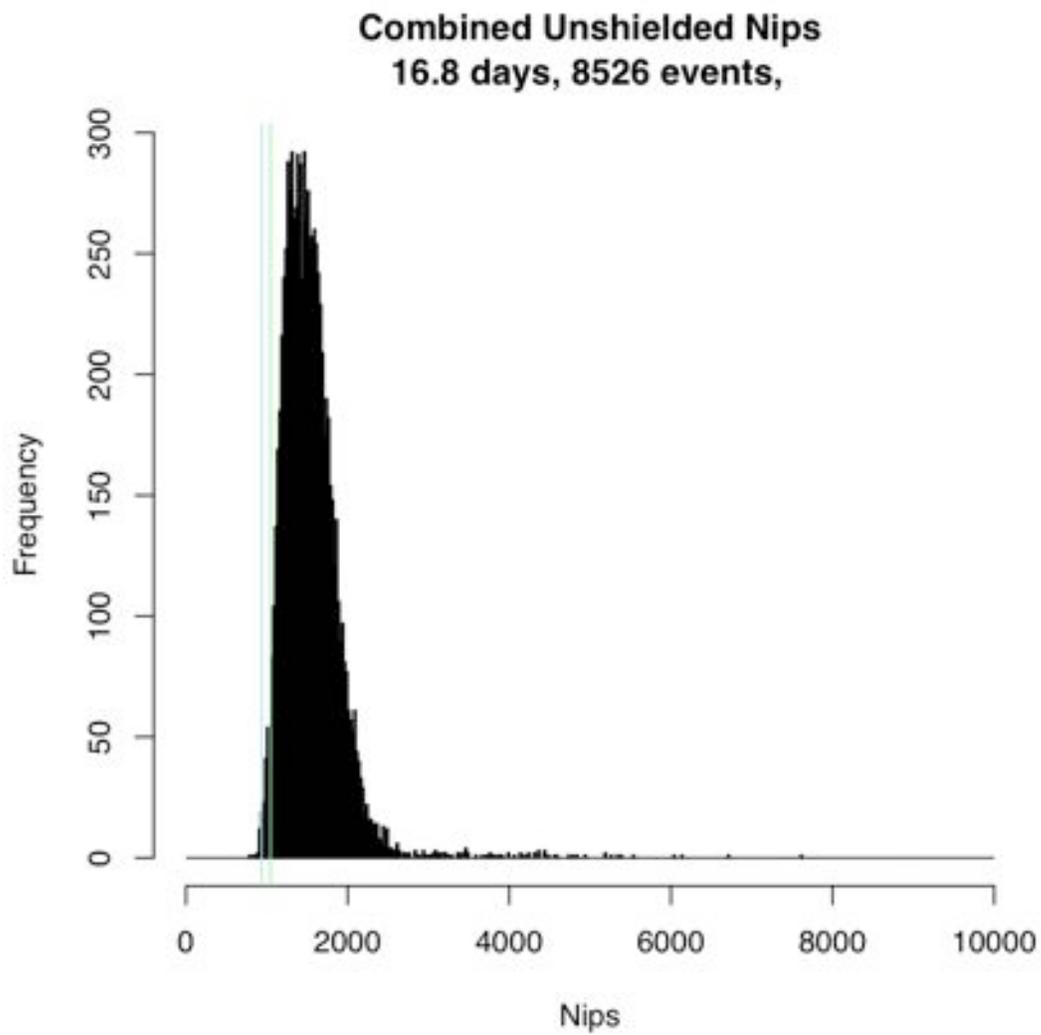

Fig. 10 – The distribution of *NIPs* for all of the unshielded data after all cuts. The green and blue vertical lines are the thresholds of the left (green) and right (blue) MWPCs during this run.

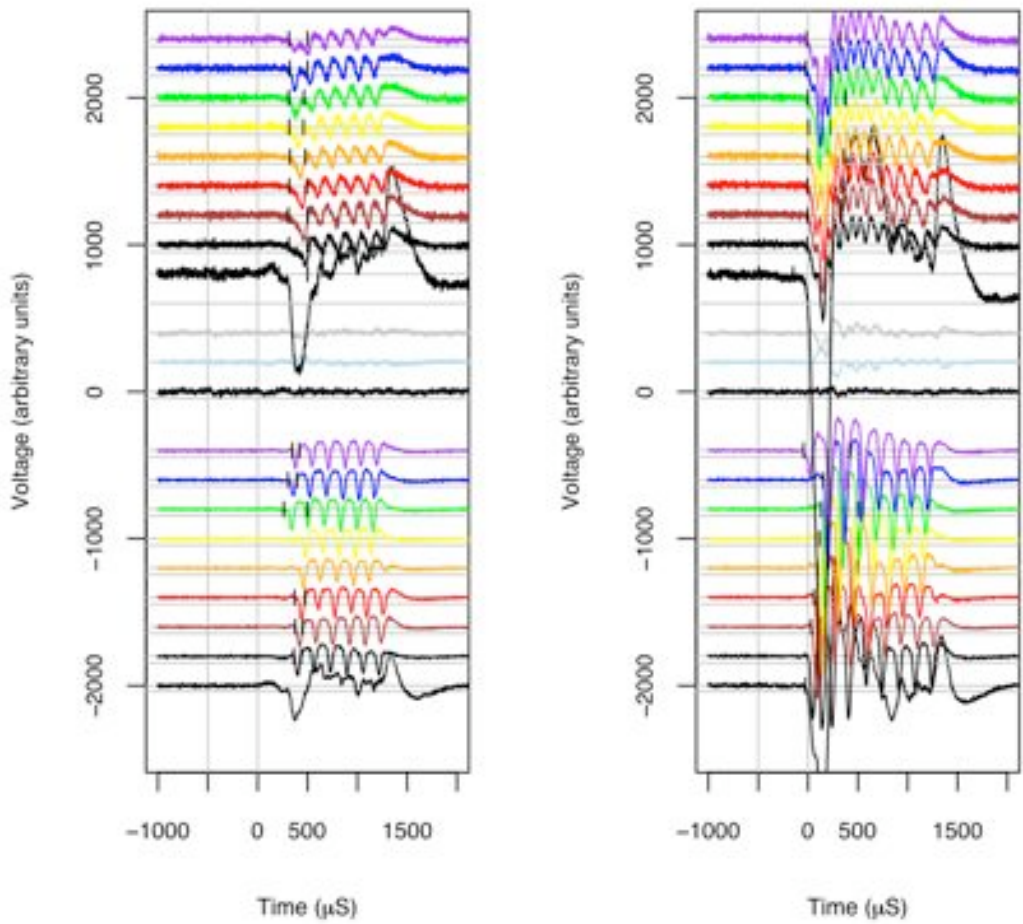

Fig. 11 - An example of an un-vetoed cathode crossing alpha (GPCC) event, an event well explained by radon decay in the gas.

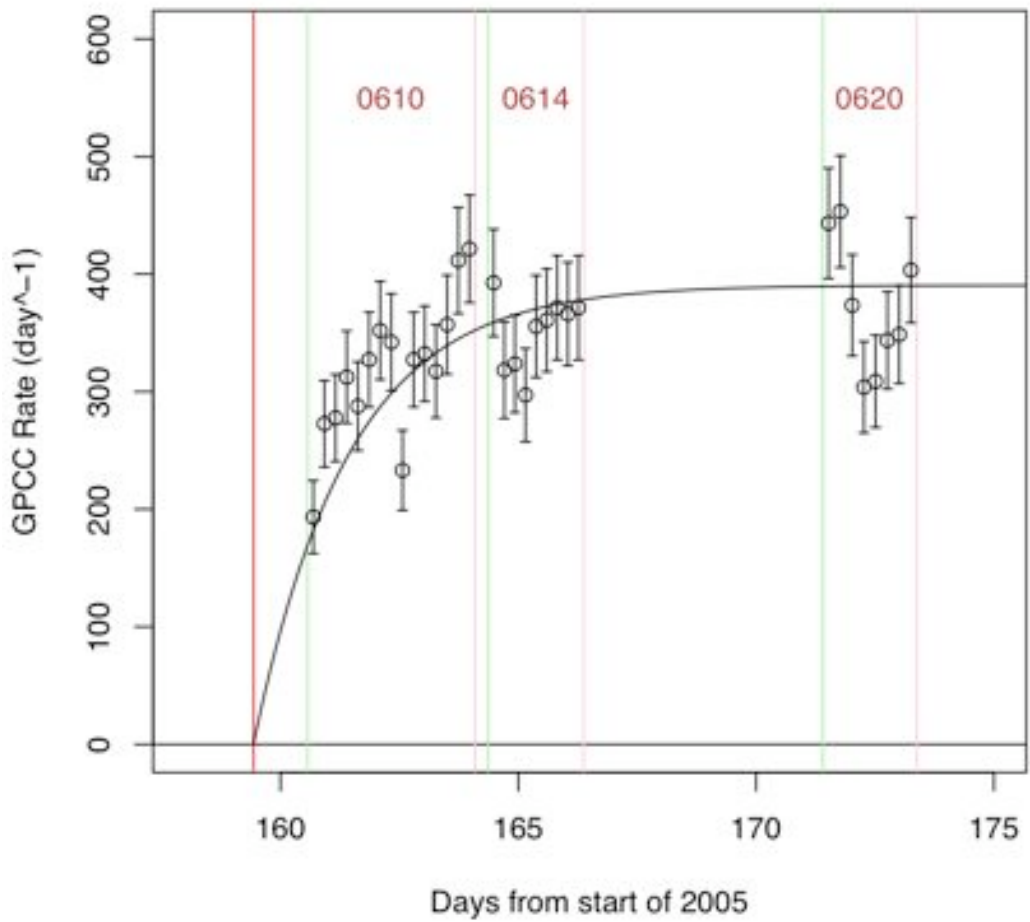

Fig. 12 – Plot showing the rate of GPCC events as a function of time for the first ~2 weeks of unshielded running. The red numbers are file designators corresponding to the date when the run started. The vertical red line shows the time when the vacuum pump was turned off and the vacuum vessel started filling with $CS_2$. Under RPR hypothesis, this would have also been the time when radon began emanating into the detector. The fit is discussed in the text and appears to be consistent with the hypothesis. The gap in the data resulted from time spent performing neutron exposures. The fit indicates $\kappa =$

1075±150 day$^{-1}$ with a reduced chi-squared of 1.5.

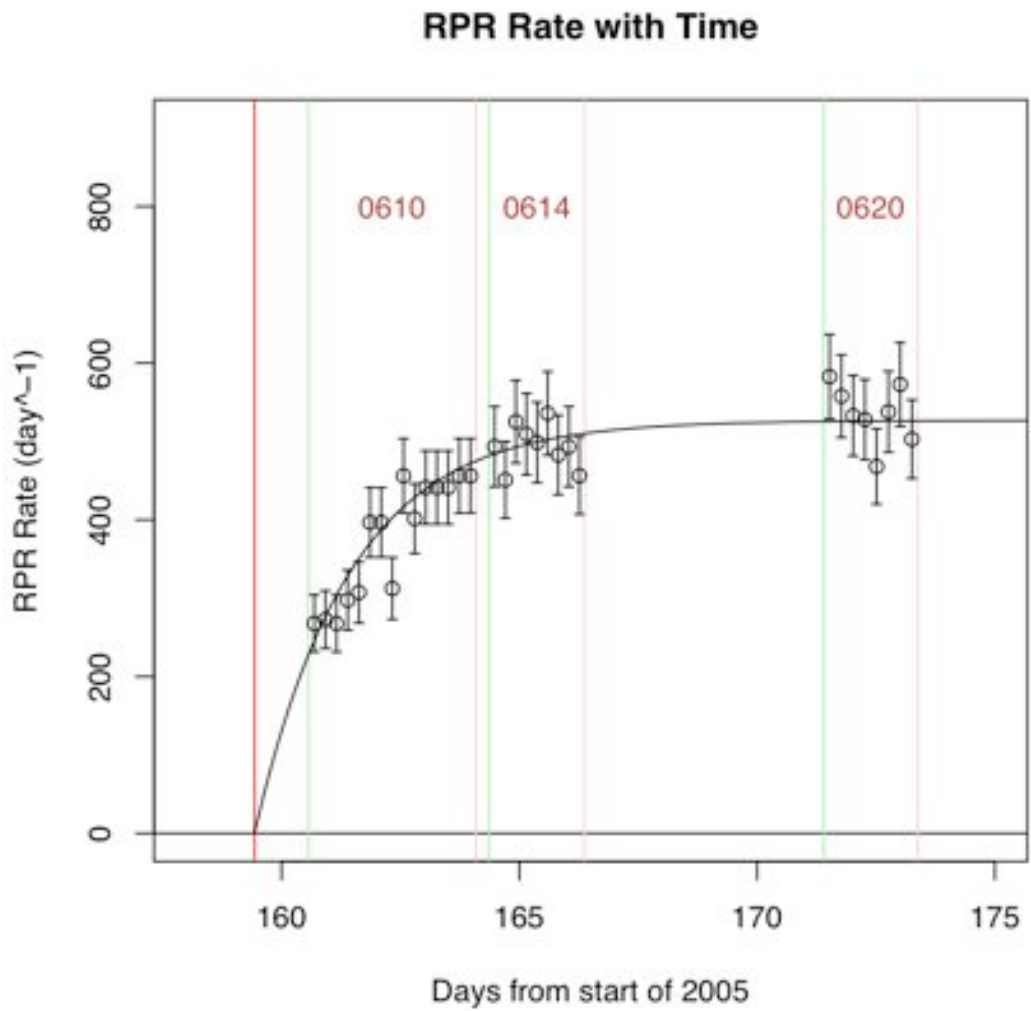

Fig. 13 – This plot shows the rate of RPR events as a function of time. The red numbers are file designators corresponding to the date when the run started. The red line shows the time at which the evacuated vacuum vessel was filled with $CS_2$. Shortly thereafter runs were started. The fit for this plot indicates κ = 1450±175 day$^{-1}$ with a reduced chi-squared of 0.55.

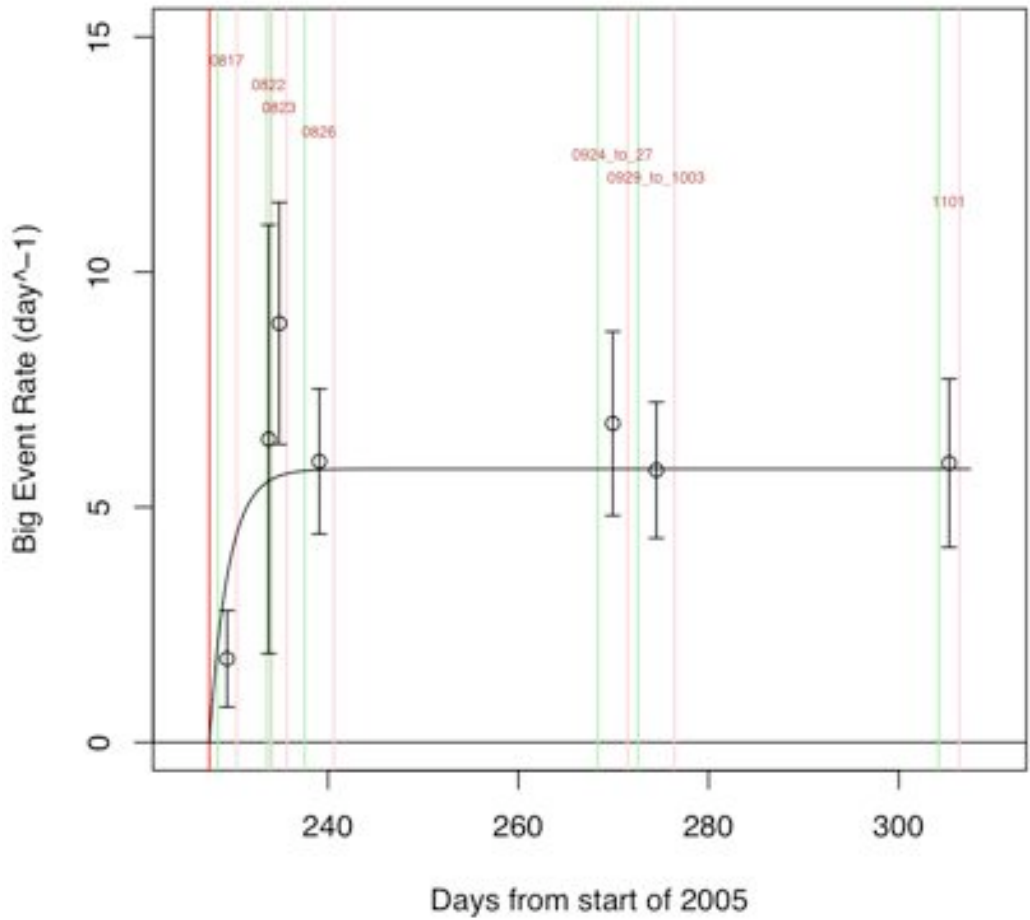

Fig. 14 – A plot of the rate of events from the shielded runs with 3000 < *NIPs* < 6000. Interestingly the rate of events in this *NIPs* window, designed to avoid RPR events shows a rate increase similar to that exhibited by the RPR events.

## References

1 R.J. Gaitskell, Ann. Rev. Nucl. Part. Sci. 54 (2004) 315.

2 J.Ellis, K.A. Olive, Y. Santoso and V.C. Spanos, Phys. Rev. D71, 095007 (2005).

3 E. Daw et al. Proc TAUP2005, Zaragoza, Spain, Sept. (2005).

4 V.A. Kudryavtsev et al., Phys. Lett. B 616 (2005) 17.

5 D.S. Akerib et al., Phys. Rev. D 73, 11102 (2006) 1.

6 V. Sanglard et al., Phys. Rev. D 71, 122002 (2005) 1.

7 E. Aprile et al., Phys. Rev. Lett. 97, 081302 (2006).

8 D.P. Snowden-Ifft, C.J. Martoff, and J.M. Burwell, Physical Review D 61 (2000) 1.

9 B. Morgan, A. Green, N. Spooner, Phys. Rev. D 103507 (2005) 781.

10 G.J. Alner et al. Nucl. Instr. & Meth. A 555 (2005) 173.

11 C.J. Martoff et al., Nucl. Instr. & Meth. A 440 (2000) 355.

12 T. Ohnuki, C.J. Martoff, and D.P. Snowden-Ifft, Nucl. Instr. & Meth. A 463 (2001) 142.

13 D.P. Snowden-Ifft et al., Nuclear Instruments and Methods section A 498 (2003) 155.

14 B. Morgan et al., Astropart. Phys. 23, (2005) 287.

15 M. Robinson et al., Nucl. Instr. & Meth. A 511 (2003) 347.

16 G.J. Alner et al. Nucl. Instr. & Meth. A 555 (2005) 173.

17 S. Agostinelli et al. Nucl. Instr. & Meth. A 506 (2003) 250. (v4.7.0.p01).

18 P.F. Smith, D. Snowden-Ifft, N.J.T. Smith, R.L. Luscher and J.D. Lewin, Astropart. Phys. 22 (2005) 409.

19 E. Tziaferi et al. Astroparticle Physics 27 (2007) 326-338.

20 M.J. Carson et al. Nucl. Instr. & Meth 546 (2005) 509-522.

21 To be published in NIMA.

22 R.J. Gaitskell, Ann. Rev. Nucl. Part. Sci. 54 (2004) 315.

23 J.P. Mclaughlin and G. Gath, Rad. Prot. Dos. 82 (1999) 257.

24 J.F. Ziegler, J.P., Biersack, U. Littmark, The Stopping and Range of Ions in Solids, vol. 1 of series Stopping and Ranges of Ions in Matter, Pergamon Press, New York (1984). http://www.srim.org/. SRIM 2003.

25 W.P. Jesse and J. Saudauskis, Physical Review 102 (1956) 389.

26 G.F. Knoll Radiation Detection and Measurement, 2$^{nd}$ edition, John Wiley and Sons, 1989.